\documentclass{elsart}
\usepackage{graphicx}
\usepackage{amsmath,amssymb}
\begin{document}

\runauthor{Case and Cherry}
\begin{frontmatter}

\title{Measurements of Compton Scattered Transition Radiation at High Lorentz Factors}
\author{Gary L. Case\corauthref{cor}}
\corauth[cor]{Corresponding author.  Email address: case@phunds.phys.lsu.edu}
\author{P. Parker Altice\thanksref{a}}
\author{Michael L. Cherry}
\author{Joachim Isbert}
\author{Donald Patterson}
\address{Dept. of Physics and Astronomy, Louisiana State University, Baton Rouge, LA 70803}
\author{John W. Mitchell}
\address{NASA Goddard Space Flight Center, Greenbelt, MD 20771}
\thanks[a]{Current address: Micron Technology, Boise, ID 83707}

\begin{abstract}
X-ray transition radiation can be used to measure the Lorentz factor of relativistic particles.  Standard transition radiation detectors (TRDs) typically incorporate thin plastic foil, foam, or fiber radiators and gas-filled x-ray detectors, and are sensitive up to $\gamma \sim 10^{4}$. To reach Lorentz factors up to $\gamma \sim 10^{5}$, thicker, denser radiators can be used, which consequently produce x-rays of harder energies ($\gtrsim 100$ keV). At these energies, scintillator detectors are more efficient in detecting the hard x-rays, and Compton scattering of the x-rays out of the path of the particle becomes important.  The Compton scattering can be utilized to separate the transition radiation from the ionization background spatially.  The use of conducting metal foils is predicted to yield enhanced signals compared to standard nonconducting plastic foils of the same dimensions.  We have designed and built an inorganic scintillator-based Compton Scatter TRD optimized for high Lorentz factors and exposed it to high energy electrons at the CERN SPS.  We present the results of the accelerator tests and comparisons to simulations, demonstrating 1) the effectiveness of the Compton Scatter TRD approach; 2) the performance of conducting aluminum foils; and 3) the ability of a TRD to measure energies approximately an order of magnitude higher than previously used in very high energy cosmic ray studies.
\end{abstract}


\end{frontmatter}

\section{Introduction}
Space-borne cosmic ray experiments require the capability to measure the energies of particles with Lorentz factors $\gamma \sim 10^{5}$ with detectors that are relatively large yet lightweight. NASA's proposed Advanced Cosmic Ray Composition Experiment for Space Science (ACCESS) mission \cite{wefel99,israel00}, for example, requires a transition radiation detector (TRD) capable of measuring the energies of cosmic rays up to 100 TeV/nucleon for particles with charge $Z>3$.  Such experiments require that the range of existing TRDs must be extended upward by an order of magnitude or more, requiring designs modified for use at these higher energies.  

Transition radiation (TR) is produced when a charged particle crosses the interface between two materials with different dielectric constants, resulting in the rapid rearrangement of the particle's electric field as it passes from one material to the next \cite{garibian58,garibian60,termik61,termik72,durand75,artru75,cherry78}. For highly relativistic particles ($\gamma = E/mc^{2} \gg 1$) the radiation is emitted at x-ray frequencies. The spectrum produced depends on the plasma frequencies and thicknesses of the two materials as well as the energy of the particle.  Typically, the materials used are a low atomic number solid such as plastic with plasma frequency $\omega_{1}$, and a gas or vacuum with plasma frequency $\omega_{2}$.  Radiation is emitted up to a frequency $\gamma\omega_1$, beyond which the spectrum is suppressed. In the usual case, where $\omega_1 \gg \omega_2$, the total intensity produced from a single interface is proportional to $Z^2\gamma\omega_1$, where $Z$ is the atomic number of the incident particle.	

The intensity from a single interface is weak.  Therefore, in practical applications, a radiator is constructed with a large number $N$ (typically $N \sim 50 - 1000$) of thin foils of thickness $l_{1}$ separated by a distance $l_{2}$ (or fiber or foam radiators with equivalent average $\left<N\right>, \left<l_1\right>,$ and $\left<l_2\right>$) with radiation produced at each of the $2N$ interfaces. Interference effects from the superposition of the amplitudes produced at each interface give rise to pronounced minima and maxima in the spectrum, with the last (highest frequency) maximum near
\begin{equation}
\omega_{\text{max}} = \frac{l_{1}\omega_{1}^{2}}{2 \pi c}(1+\rho) \;,
\label{eq:omega_max}
\end{equation}
where $\rho$ is 1 for a metal and 0 for a nonconductor.   As the particle energy increases, the total radiated intensity increases up to a Lorentz factor
\begin{equation}
\gamma_{\text{s}} \approx \frac{0.6 \omega_{1}}{c} \sqrt{l_{1}l_{2}(1+\rho)} \;,
\label{eq:saturation}
\end{equation} 
above which saturation sets in due to the interference.  We have included here the possibility of a nonzero conductivity which introduces an imaginary part to the wave vector and leads to an effective plasma frequency $\omega_{1}\sqrt{(1+\rho)}$ \cite{cherry02}. The saturation energy and characteristic frequency can be tuned by varying the radiator foil material, thickness, and separation.

An x-ray detector appropriate for absorbing the TR x-rays must be placed after the radiator.  The radiation is emitted at an angle $\theta \sim 1/\gamma$ with respect to the incident particle direction, so for high Lorentz factors the x-rays are spatially inseparable from the ionization energy deposited in the detector by the particle itself.  Therefore, in conventional applications, the detector must be made thin in order to minimize the ionization signal, yet with sufficient stopping power to absorb the x-rays.  For $\omega_{\text{max}}$ less than about 40 keV, gaseous detectors (e.g. Xenon-filled wire chambers) are typically employed.  In order to improve statistics and provide redundancy, a complete TRD consists of multiple layers of radiators and x-ray detectors.  

Such TRDs have been used successfully both at accelerators and in space. In most cases, the TRD is employed as a threshold device to identify particle types: For example, a meson or hadron may be accompanied by a small TR signal, while an electron of the same energy but larger $\gamma$ is characterized by a large signal. In the case of the Space Shuttle CRN experiment \cite{grunsfeld88,swordy90}, a fiber TRD was used to measure the energies of cosmic ray nuclei with $\gamma \ge 3\times10^{3}$.  Ref.\ \cite{favuzzi} gives an extensive review of TR applications and radiator configurations.  A brief listing of cosmic ray experiments incorporating TRDs is given in \cite{giglietto}.

In order to increase the maximum particle energy $\gamma_{\text{s}}$, one must increase the plasma frequency (or equivalently, density), thickness, and/or spacing of the foils (Eq.~\ref{eq:saturation}). In a space instrument, the overall thickness will be constrained, putting a limit on $Nl_{2}$ (assuming $l_{2} \gg l_{1}$).  Increasing $\omega_{1}$ by using metal foils instead of plastic, for example, and/or $l_{1}$ results in a hardening of the x-ray spectrum produced (Eq.~\ref{eq:omega_max}). Metal foils have been used in early accelerator tests \cite{Yuan}, and in particular lithium foils have been used in order to minimize the absorption at low x-ray frequencies \cite{fischer,cobb}. In the case of very high energies, though, with $\gamma_{\text{s}} \approx 10^{5}$ and a typical spacing $l_{2} = 0.1-1$ cm, $\omega_{\text{max}} \approx 0.4 \gamma_{\text{s}}^{2}c/l_{2}$ can be in excess of several hundred keV.  Gas detectors are then no longer efficient in detecting these hard x-rays.  Although \cite{wakely02} describes the use of gas detectors near $\gamma_{\text{s}} \approx 10^{5}$ by optimizing the radiator design, scintillators such as NaI or CsI provide an efficient alternative at these Lorentz factors and corresponding high x-ray energies. The higher density of the scintillators leads to an increase in the ionization energy deposited by the particle as it traverses the detector.  However, as the TR spectrum hardens, Compton scattering in the radiators becomes important, becoming the dominant photon interaction above $\approx 40$ keV.  A significant portion of the x-rays produced are scattered out of the path of the incident particle. Thus, a detector that is segmented or positioned outside of the beam can efficiently detect the TR signal spatially separated from the ionization.  

We describe here the test of a scintillator-based Compton Scatter TRD for high Lorentz factor particles, including the use of metal foils, based on the results of accelerator measurements with high energy electrons at the CERN SPS.  Comparisons of the measured results with detailed simulations will also be presented.  

\section{Experimental setup}
A scintillator-based Compton Scatter TRD was designed to investigate the predicted increase in saturation energy obtained by using thick, dense radiator materials including Mylar ($\rho = 1.4$ g/cm$^{3}$), Teflon ($\rho = 2.0$ g/cm$^{3}$), and aluminum ($\rho = 2.7$ g/cm$^{3}$).  Metal foils are of particular interest because of the characteristic enhancement in the signal expected due to the nonzero conductivity \cite{cherry02}.  

\begin{figure}[b]
\begin{center}
\includegraphics[width=3.4in]{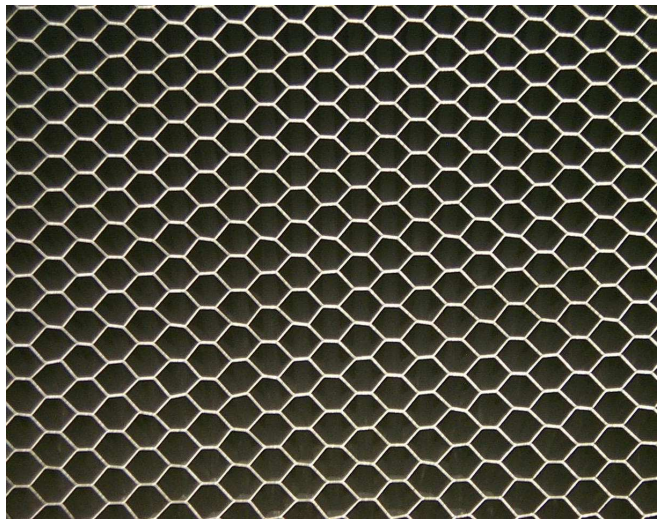}
%\vspace{-0.5pc}
\caption{Aluminum honeycomb structure. The particle beam entered the radiator at the top and moved vertically downward through the structure. \label{fig:honeycomb}}
\end{center}
\end{figure}

\begin{table*}[t]
\begin{center}
\caption{Parameters of radiator configurations tested\label{tab:rad_par}}
\begin{tabular}{lcccccc}
\hline
Radiator & $\omega_{1}$ & $l_{1}$  & $l_{2}$  & N & $\omega_{\text{max}}$ & $\gamma_{\text{s}}$ \\
         &    (eV)      & ($\mu$m) &  (mm)    &   &       (keV)           &                  \\
\hline
Thin Mylar         & 24.4 & 122 & 3.4 & 50 & 61  & $4.9\times 10^{4}$ \\
Thick Mylar        & 24.4 & 254 & 3.4 & 50 & 122 & $6.9\times 10^{4}$ \\
Thin Teflon        & 28.5 & 122 & 3.5 & 50 & 83  & $5.8\times 10^{4}$ \\
Thick Teflon       & 28.5 & 203 & 3.4 & 50 & 133 & $7.2\times 10^{4}$ \\
Aluminum           & 32.7 & 133 & 3.8 & 48 & 230 & $9.9\times 10^{4}$ \\
\hline
\end{tabular}
\end{center}
\end{table*}

For the plastics, radiators of $N=50$ foils were constructed by attaching 19.1 cm $\times$ 18.4 cm plastic foils to 3.4 mm thick wood frames and stacking them together. For the aluminum, each radiator consisted of seven 2.7 cm thick honeycomb panels bundled together and aligned with the cells perpendicular to the particle beam.  The honeycomb was a composite material chosen both for its dimensions and its adaptability as a combined detector-plus-structure for a space instrument (Fig.\ \ref{fig:honeycomb}).  Particles passing through the structure  passed through either 1) a section of foils perpendicular to the beam in which two 3 mil sheets glued together form a foil with an effective $l_{1}=6$ mil, $l_{2} = 5.2$ mm, and $N=35$ foils along the particle trajectory or 2) a section of foils at a $41^{\circ}$ angle with respect to the particle beam, resulting in an effective foil thickness $l_{1}=3 {\rm \;mil}/\sin 41^{\circ} = 4.6$ mil, $l_{2}$ ranging from 0 to 5.2 mm, and $N=70$.  The yield from a composite material (e.g., a foam) with average values $\left<l_{1} \right>$, $\left<l_{2}\right>$, and $\left<N\right>$ has been shown to be essentially the same as from a regular foil radiator with the same $l_{1}$, $l_{2}$, and $N$ \cite{prince75}.  We therefore calculate the Al honeycomb effective parameters as averages of configuration 1 weighted by 46\% (to account for the fraction of the area perpendicular to the beam covered by configuration 1) and configuration 2 weighted by 54\%.  The resulting average honeycomb radiator parameters are given in Table~\ref{tab:rad_par}, along with the parameters of the plastic radiator configurations tested.  The total length of each radiator was 19 cm.

Each radiator was viewed by three x-ray detectors, each consisting of a 19 cm $\times$ 19 cm $\times$ 5 mm thick NaI(Tl) crystal hermetically sealed between a 6.4 mm thick glass optical window on one flat face and a 0.75 mm thick aluminum entrance window on the other face. An ultraviolet transmitting Lucite lightguide was coupled to the glass window using optical grease, reducing the aperture to a 13 cm diameter circle. An Electron Tubes 9390KB 130 mm photomultiplier tube with a standard bialkali photocathode was mated to the lightguide with optical grease. The lightguide was wrapped in aluminum foil and the whole assembly wrapped with black tape to make it light tight.

\begin{figure}[t]
\begin{center}
\includegraphics[width=4.5in]{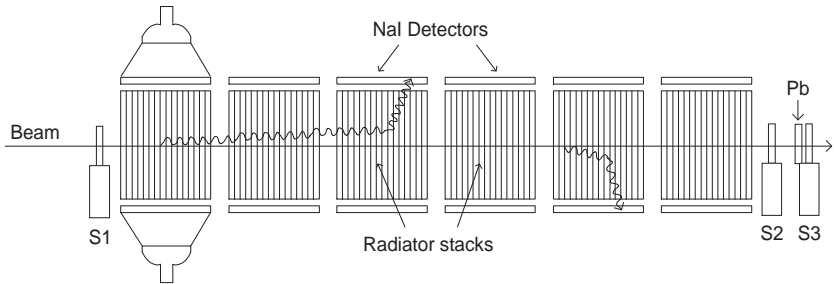}
\caption{Schematic of the experimental setup, as seen from above.  Lightguide/PMT assemblies are shown for the first module only.  An additional NaI detector (not shown) was positioned above each radiator.  The location of the beam definition scintillators S1 and S2 and shower counter S3 are also shown.  TR x-rays are produced in the forward direction and can Compton scatter out of the beam and into the NaI.\label{fig:schematic}}
\end{center}
\end{figure}

Six identical modules were constructed, with each module containing a radiator and the three NaI(Tl) detector assemblies, one on each side of the radiator and one above the radiator outside of and parallel to the beam (Fig.~\ref{fig:schematic}).  The modules were positioned one behind the other along the beam and aligned such that the particle beam travelled down the center of the modules.  Only x-rays scattered at large angles away from the beam were then detected.

For the first accelerator run, the 18 PMT signals were fed into CAMAC-based, 11-bit CAEN C205A charge ADCs and read out with the CERN CMS H2A DAQ computer. For the second accelerator run, the signals from the PMTs were fed into custom-built 8-channel front end modules which contained a charge integrator, peak detect and hold circuit, and gain-adjustable amplifier. Analog-to-digital conversion was performed with a 64-channel, 12-bit National Instruments PCI-6071E DAQ board running in a PC and controlled under LabVIEW 6i.

The instrument was exposed to high energy electrons at the CERN SPS H2A test beam site in August/September 1999 and again in August/September 2001.  Beam energies ranged from 7 to 150 GeV, covering the range of Lorentz factors $\gamma = 1.4\times 10^{4}-2.9\times 10^{5}$.  A set of scintillators in the beam upstream of the TRD provided event triggering. The trigger rate was kept to about 1 kHz to avoid deadtime in the DAQ system. Beam definition scintillators in front of and behind the TRD (S1 and S2 in Fig.~\ref{fig:schematic}) flagged events for which the electrons showered within the radiator stacks.  A Pb shower counter (S3) was placed downstream of S2 to flag pions present as a contaminant in the higher energy beams.  

Energy calibration runs were performed both immediately before and after the beam runs using radioactive $^{133}$Ba (81, 303 and 356 keV) and $^{137}$Cs (662 keV) x-ray sources. In order to account for bremsstrahlung and other background produced by the electrons in passing through the radiators and upstream material, a background run was performed for each radiator configuration in which the radiators were replaced by solid blocks with the same material and thickness (in g/cm$^{2}$) as the radiators.  
 
\begin{figure}[t]
\begin{center}
\includegraphics[width=3.4in]{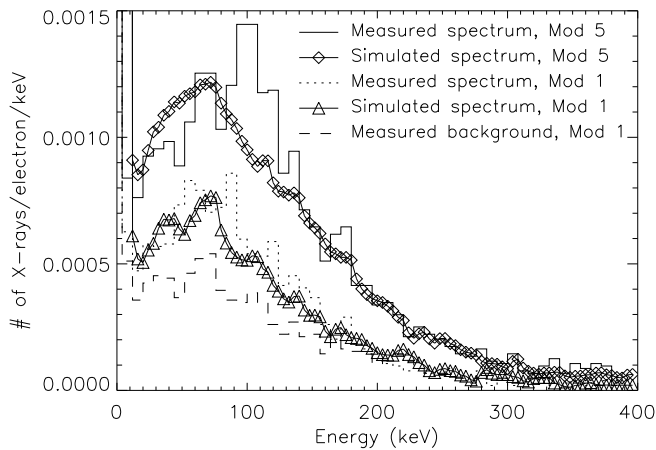}
\caption{Spectra measured using 150 GeV electrons with aluminum honeycomb radiators and solid aluminum background plates. The lowest histogram (dashed line) shows the measured background spectrum in Module 1. The middle histogram (dotted line) shows the spectrum (TR plus background) measured in Module 1 with the Al foils; triangles (\footnotesize$\triangle$\small) represent the calculated spectrum for Module 1. The upper histogram (solid line) and diamonds (\normalsize$\diamond$\small) show the measured and calculated spectra, respectively, in Module 5.\label{fig:spectra}}
\end{center}
\end{figure}

\section{Results}
For each material, a background run and a foil run were made for each electron energy used. 
Figure~\ref{fig:spectra} shows spectra obtained  for aluminum foil and background runs at 150 GeV. 
Several conclusions can be drawn immediately: first, Compton scattered transition radiation is being detected away from the path of the incident electron, at levels well above that of the background; second, the detected TR x-ray spectrum peaks near 100 keV, with some x-rays detected at energies $> 200$ keV; and third, the measured intensity increases as the particle moves downstream through the set of radiator/detector modules. 

Calculated spectra are produced by a Monte Carlo routine based on the description in \cite{cherry02}, in which the differential intensity per unit solid angle per unit frequency is expressed in terms of the coherent sum of the complex amplitudes from the individual interfaces \cite{termik72,artru75,cherry78}. The effect of the metal foils is included by incorporating an effective absorption cross section (i.e., the imaginary part of the wave vector) that depends on the foil conductivity. Individual x-rays in the range $2 - 1000$ keV are produced at random locations along the trajectory and followed through the geometry of the radiator stacks and detector modules taking into account the effects of photoelectric absorption, Compton scattering, fluorescence, and escape in both the detectors and radiators, and photoelectron statistics and electronic resolution in the scintillators, photomultipliers, and electronic readout. Examples of the pure TR spectra, showing the characteristic interference pattern and a maximum in the predicted spectrum near 200 keV, are shown in \cite{cherry02}. The calculated TR plus background spectra are shown here: The triangles in Fig.~\ref{fig:spectra} show the result of convolving the measured background in Module 1 with the expected TR signal; the diamonds show the result in Module 5. The predicted signals agree well with the measured TR-plus-background spectra.

\begin{figure}[t]
\begin{center}
\includegraphics[width=3.4in]{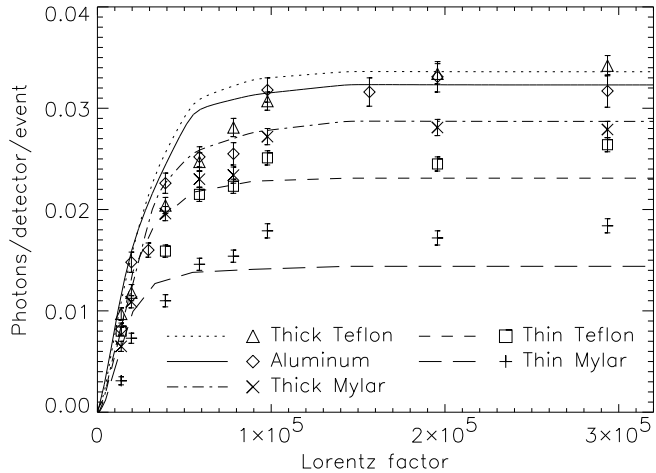}
\caption{Average number of photons detected in the energy range $35-500$ keV per detector per event in the first module as a function of electron energy for various radiator configurations.  The error bars represent statistical errors.   \label{fig:saturation}}
\end{center}
\end{figure}

Figure~\ref{fig:saturation} shows the total number of photons detected per NaI detector summed over the x-ray energy range $35-500$ keV as a function of electron energy.  The points show the measured data; the curves show the results calculated as described above. The observed saturation Lorentz factors for the thick Teflon and aluminum honeycomb are $\approx 10^{5}$, as expected from the calculated values in Table~\ref{tab:rad_par}.  The calculations reproduce the differences in detected yield between different radiator materials, the effects due to different thicknesses of the same material, and the dependence on electron energy to an accuracy of $\leq 20\%$.  

As shown in Fig.~\ref{fig:spectra}, the number of x-rays measured in each detector depends on the position of the module.  Most of the TR x-rays produced in the beginning of the first module will pass through that module without interacting ($73\%$ probability in an Al honeycomb radiator at 100 keV).  But as they encounter the radiator material in successive modules, they can Compton scatter and be absorbed in a downstream detector.  For the Al honeycomb, there is an 80\% probability that a 100 keV x-ray created at the front of the first module will Compton scatter in the radiators before leaving the last module.  This feedthrough effect enhances the number of photons detected in modules downstream and can be used to advantage in designing a practical detector \cite{cherry02}.  
For 150 GeV electrons and Al radiators over the x-ray energy range $35-500$ keV, the ratios of x-rays detected in the downstream modules compared to Module 1 are $1.97 \pm 0.12$, $3.34 \pm 0.19$, $3.63 \pm 0.21$, $4.00 \pm 0.22$, and $4.31 \pm 0.25$ for Modules 2--6, respectively.  The corresponding predicted ratios are $2.31 \pm 0.04$, $3.30 \pm 0.05$, $3.95 \pm 0.05$, $4.24 \pm 0.04$, and $3.74 \pm 0.13$. 

\section{Conclusion}
A new Compton Scatter Transition Radiation Detector capable of measuring the x-rays produced by particles with Lorentz factors near $\gamma = 10^{5}$ has been built and successfully tested.  Compton scattered TR x-rays were detected outside of the particle beam using relatively thick NaI scintillator detectors, effectively isolating the TR signal from the ionization signal.  For thick Teflon and Al honeycomb radiators, the detected x-ray spectrum peaks near 100 keV with some x-rays of energy $> 200$ keV detected, and saturation Lorentz factors near $10^{5}$ were achieved.  The detected yields for most radiator configurations agree with detailed simulations, including the enhancement expected from metal foils.  The measurements tend to saturate at slightly higher particle energies than predicted, and the measured number of photons at saturation are $\sim 10-20\%$ higher than predicted for the thin Mylar and Teflon radiators.  These discrepancies are presumably due to nonuniformities known to be present in the radiator material.  Likewise, the measured peak of the TR plus background spectrum from the aluminum honeycomb radiator occurs approximately 30 keV higher than expected.  This discrepancy presumably reflects the approximations inherent in treating an irregular honeycomb structure as a periodic stack of foils.  In the case of hard x-ray energies, photon feedthrough enhances the signal in the downstream detectors, again as predicted by the simulations (within $\sim 15\%$).  

\section{Acknowledgments}  
This work was supported by NASA grant NAG5-5177 and NASA/Louisiana Board of Regents grant NASA/LEQSF-IMP-02.  The authors wish to thank A. Aranas, S. Apewokin, T. Brown and O. Shertukde for their many hours of work during the construction and testing of the radiators and detectors; the staff at CERN for their superb cooperation and assistance, particularly D. Lazic and G. Bencze; and J. Anderson, J. Marsh and C. Welch for assisting with the data analysis.

\end{document}